\pdfoutput=1
\documentclass[%
 aip,  
 longbibliography,
 amsmath,amssymb,
 reprint,%
 floatfix,
]{revtex4-1}

\usepackage[dvipdfmx]{graphicx}
\usepackage{bmpsize}
\usepackage{dcolumn}
\usepackage{bm}

\usepackage[utf8]{inputenc}
\usepackage[T1]{fontenc}
\usepackage{mathptmx}
\usepackage{float}
\usepackage{xcolor}

\usepackage{natbib}
\usepackage{hyperref}
\usepackage{microtype}

\usepackage{silence}
\begin{document}


\title{Impact of the nucleation of charge clusters on the retention of memristors: a self-consistent phase field computational study }

\author{Foroozan S. Koushan}%
 \email{foroozan.koushan@gmail.com.}
\author{Nobuhiko P. Kobayashi}%
\affiliation{\leftskip=35pt \rightskip=35pt {Nanostructured Energy Conversion Technology and Research (NECTAR), Univ. of California Santa Cruz, Santa Cruz, CA 95064, USA}}%
\affiliation{\leftskip=35pt \rightskip=35pt {Department of Electrical and Computer Engineering, Jack Baskin School of Engineering, University of California, Santa Cruz, CA 95064 USA}}%


\begin{abstract}
\leftskip=35pt \rightskip=35pt {In recent years, resistive RAM often referred to as memristor is actively pursued as a replacement for non-volatile-flash memory due to its superior characteristics such as high density, scalability, low power operation, high endurance, and fast operating speed. However, one of the challenges that need to be overcome is the loss of retention for both ON- and OFF-states – the retention loss. While various models are proposed to explain the retention loss in memristors consisting of a switching layer, in this paper, we propose that the nucleation of clusters made of electrical charges – charge-clusters – in the switching layer acts as a potential root cause for the retention loss. The nucleation results from localized electric-field produced intermittently during cyclic switching operations. We use the phase-field method to illustrate how the nucleation of charge- clusters gives rise to the retention loss. Our results suggest that the degree at which the retention loss arises is linked to the number of cyclic switching operations since the probability at which nucleation centers form increases with the number of cycle switching operations, which is consistent with a range of experimental findings previously reported.}
\end{abstract}

\maketitle

\section{\label{sec:level1}Introduction:}

The emergence of novel resistive memories such as resistive RAM (RRAM)\cite{t1}, phase change memory (PCM) \cite{t2}, and spin transfer torque RAM (STTRAM)\cite{t3}, aiming at replacing or complementing flash memories and other silicon-based memories including dynamic random-access memory and static random-access memory, offers many advantages in terms of size, speed, and scaling. Among these resistive memories, filamentary RRAM deems to be most promising from the perspective of its structural simplicity, ease of integration at the back-end-of-line (BEOL), scalability, and fast switching speed\cite{t4,t5}. However, for a successful implementation of the technology, various challenges such as variabilities of electrical properties at the device-to-device and cycle-to-cycle level\cite{t7}, random telegraph noise\cite{t8}, and loss of data\cite{t9} must be resolved. Of these challenges, the loss of data refers to the loss of the ability of retaining low-resistance-state (i.e., ON-state) and/or high-resistance-state (i.e., OFF-state) – the retention loss.

The retention of data can be seen as a trade-off\cite{t6}, that is, an improvement in the retention of data often results in a degradation of endurance and vice versa. Furthermore, the retention of data is found to be adversely interfered with changes in temperature, application of SET/RESET biases, and structural alternations occurring at the bottom and top electrode interfaces\cite{t10}. The retention loss that occurs over a period of time shorter than 1 minute is attributed to statistical fluctuations of electrical conductance\cite{t11}, whereas the retention loss that takes place over a period of time longer than 1 minute is often modeled based on the diffusion of oxygen vacancies\cite{t29}. 

Regardless of underlying physics, the retention loss results when resistance of OFF-state, $R_{OFF}$, decreases over time and/or resistance of ON-state, $R_{ON}$, becomes unstable and decreases over time, even when the device is no longer under external electrical bias. Additionally, reversible transitions between OFF-state and ON-state – cyclic switching operations – are often interpreted as the formation and annihilation of electrically conducting filaments (ECFs)\cite{t31,t32,t19}; thus, one possible scheme for the retention loss can be illustrated by allowing a dielectric film – switching layer – responsible for the resistive switching to evolve structurally away from those that define OFF-state and ON-state. Such structural evolutions in the switching layer with or without the presence of ECFs are delineated by, for instance, introducing the nucleation of clusters made of electrical charges – charge-clusters– which is described in the context of the formation of metallic nuclei that can occur in the switching layer under the influence of electric-field and lead to such a phase transition as the insulator-metal transition\cite{t15}. Furthermore, studies have shown that even when the metallic phase is energetically unfavorable, the insulator-metal transition can still take place with electric-field being sufficiently high\cite{t13}, leading to the development of models based on the formation of electrically conducting nuclei in describing switching mechanisms for PCM and ferroelectric memory\cite{t30}.

Since the nucleation of charge-clusters under the influence of electric-field appears to be responsible for the insulator-metal transition regardless of the underlying microscopic mechanisms (e.g., densification, crystallization, electron solvation)\cite{t16}, it is sensible to extend this view to filamentary RRAM and memristors in which electrically insulating regions, in the switching layer, experience localized electric-field during cyclic switching operations\cite{t17}. In RRAM, relative magnitude of chemical potential of insulating, unstable conductive, and metastable conductive phases and their corresponding thermodynamic energy barriers determine the nucleation and growth of charge-clusters, and thus, their effects on OFF-state and ON-state\cite{t20}. For instance, emerging charge-clusters in the switching layer can potentially modify fragments of ECFs present in OFF-state and resulting in reduction of $R_{OFF}$; they can also possibly connect multiple ECFs present in the switching layer in ON-state and push $R_{ON}$ to even lower resistance and cause a failure during an erase operation. Furthermore, since the number of charges in a switching layer is conserved, the emergence of charge-clusters can also fracture ECFs, resulting in the retention loss of ON-state. 

Given all these possible scenarios exhibited by the formation of charge-clusters, in this paper, we incorporate the nucleation and growth of charge-clusters in a switching layer of RRAM initially set to either ON-state or OFF-state by leveraging our previously developed approach based on the phase-field method in illustrating the formation and annihilation of ECFs\cite{t19} in order to describe the retention loss of RRAM. Our results suggest that even if charge-clusters do not contribute to cyclic switching operations of RRAM, they can potentially play a crucial role in setting a root cause of the retention loss for both OFF-state and ON-state. 

\section{\label{sec:level2}NUCLEATION AND CAHN-HILLIARD MODEL}

The classical nucleation theory developed by Becker and Döring\cite{t21} is based on a thermodynamic approach by which the Gibbs free energy of a system is minimized, using energies associated with macroscopic entities such as surface to develop expressions for the rate of nucleation. This thermodynamic approach is extended to various types of phase transitions\cite{t22}, being established as a traditional way of describing the crystallization of solid. In general, there are two types of nucleation: the nucleation that occurs at nucleation sites located on solid surfaces contacting liquid or vapor is referred to as heterogeneous nucleation. In contrast, homogeneous nucleation occurs spontaneously and randomly in a host phase brought to a supercritical state such as a supersaturation, which is relevant to our study. The presence of external electric-field influences the homogeneous nucleation, and the rate of the nucleation depends on the ratio of the dielectric constant of solution and that of solid\cite{t23}. Several experimental studies clarified that the presence of electric-field influences nucleation processes \cite{t24,t25}, highlighting the existence of critical electric-field above which nucleation was promoted. For example, electric-field in the range of 0.1 to 1 $MV/cm$ was found to discernibly increase the rate of nucleation of ice\cite{t26}. Given these findings, our premise is that the homogeneous nucleation of charge-clusters occurs as a consequence of the presence of local electric-field higher than the average electric-field defined globally by the applied electric potential across a switching layer in a memristor. Consequently, the likelihood of the formation of charge-clusters increases as a switching layer undergoes increasing number of switching cycles throughout its lifetime (i.e., the total time over which a switching layer is stressed by electric-field). While the classical nucleation theory describes the nucleation, the growth of nuclei is expressed as the evolution of size distribution of nuclei over time\cite{t28}. 
\\
In general, nucleation energy which is equivalent to the change in the total free energy ${\Delta}F$ associated with the formation of a nucleus with radius $r$ in a homogeneous system is composed of a term representing a reduction in the free energy due to the generation of a spherical nucleus $V{\Delta}f_{(c)}$ and a term due to an increase in the interfacial energy $A$${\gamma}$:

\begin{eqnarray}
\Delta F= f_{n} = -V \Delta f_{(c)}+A \gamma
=-\frac{4}{3} \pi r^3 \Delta f_{(c)}+4 \pi r^2 \gamma
\label{eq:nuc_one}
\end{eqnarray}
where $V$ is volume and $A$ is the area of nucli center, $\gamma$ is the surface energy, $\Delta f_{(c)}$ is the free energy associated with the volume of the nucli site, and $c_{(r,t)}$ is the composition of the nucleus.

In an inhomogeneous system (e.g., a switching layer with ECFs), the charge concentration $c_{(r,t)}$, where $r$ is a position vector and $t$ is time, in a switching layer is a field and not a scalar variable, thus, the Cahn-Hilliard model adds a correction to its homogeneous free energy function to account for spatial inhomogeneity:
\begin{eqnarray}
F=\int_{S}[f_n(c_{(r,t)})+\frac{1}{2}\kappa |\nabla c_{(r,t)}|^2]ds
\label{eq:nuc_two}
\end{eqnarray}
where $f_(c_{(r,t)})$ is the homogeneous free energy, $\kappa$ is the gradient energy coefficient, and  $\frac{1}{2}\kappa |\nabla c_{(r,t)}|^2$ is the gradient energy that provides the first-order correction for the inhomogeneity, allowing interfacial energy to be modeled in the phase-field method; above integration is done over the entire system represented by $S$.  Since the concentration is a conserved quantity, the dynamical evolution of $c_{(r,t)}$ is obtained by invoking a general form of phase-field conservation law, as the first-order variation of the free-energy functional equation which eventually results in: 
\begin{eqnarray}
\frac{\partial c_{({\bf r},t)}}{\partial t}= \nabla . M  \nabla \left[ \frac{\partial {f_n(c_{({\bf r},t)})}}{\partial c_{({\bf r},t)}} - \nabla . \kappa \nabla c_{({\bf r},t)}\right]
\label{eq:nuc_three}
\end{eqnarray}

\begin{figure}[ht]
	\centering
	\includegraphics[width=0.45\textwidth]{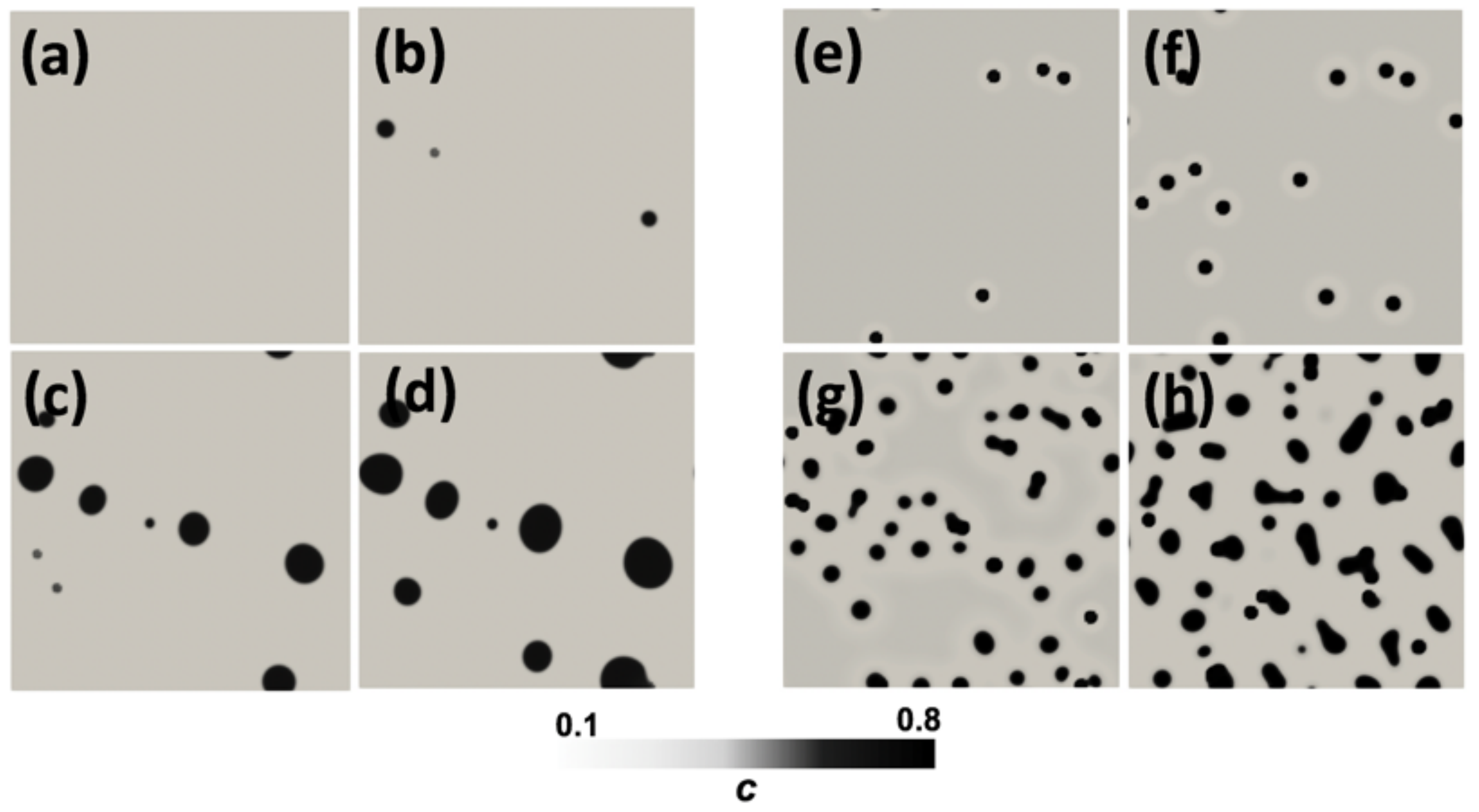}
	\caption{The nucleation and growth of clusters in a 500 $nm$ x 500 $nm$ dielectric film over time. The periodic boundary conditions are assumed on left, right, top and bottom of the system. Panels (a) through (d) are for $P_{n}$ = 1x10-7$c_{(\bf r,t)}$; whereas panels (e) through (h) are for $P_{n}$ = 5x10-6$c_{(\bf r,t)}$. Nucleation and growth of charge-clusters was captured at a specific time interval for the two cases: (a) and (e) t = 25, (b) and (f) t =75, (c) and (g) t =100, (d) and (h) t =150 unit time.}
	\label{fig:nuc_one}
\end{figure}

Where $M$ is the mobility of the conserved variable representing a physical property of the system and is assumed to be constant. In the use of the phase-field method, Eq.~(\ref{eq:nuc_three}) was applied to a system of 500 $nm$ x 500 $nm$ – much larger than the dimensions of a switching layer described later – to illustrate the nucleation and capture the growth of charge-clusters over time as shown in Fig.~\ref{fig:nuc_one}. Within the system, charges were initially distributed at $c_{(\bf r,t)}$ randomly chosen in the range of 0~0.3 (Note: $c_{(\bf r,t)}$ is forced to take a value in the range between 0 to 1; thus it represent a relative charge density) and spontaneous nucleation is driven by the presence of variations in $c_{(\bf r,t)}$, through the concentration dependent probability function, in addition, the periodic boundary conditions were imposed for the top and bottom, and the left and right sides of the system in Fig.~\ref{fig:nuc_one}. In the classical nucleation theory, a nucleation rate $R$ in the current context is defined as: 
\begin{eqnarray}
R=P_{n} \exp \left(\frac {-\Delta f^*_{n}}{K_{B}T}\right)
\label{eq:nuc_four}
\end{eqnarray}
where $\Delta f^*_{n}$ is the critical free energy of nucleation at which an nucleus is stabilized, $P_{n}$ is the nucleation probability function which depends on the mass and density of charge-clusters associated with the interaction with a dielectric matrix\cite{t33}, $k_{B}$ is the Boltzmann constant, and $T$ is the temperature. 

Studies indicate that supersaturation and charge density are two parameters that have dominant impacts on $P_{n}$\cite{t32,t33,t34}, thus, we assert that $P_{n}$ linearly depends on $c_{(\bf r,t)}$. Fig.~\ref{fig:nuc_one}(a)-(d) and (e)-(h) display two examples obtained by using two different Pn: $P_{n}$ = 1x10$^-7$$c_{(\bf r,t)}$ and $P_{n}$ = 5x10$^-6$$c_{(\bf r,t)}$, respectively. In each of the two cases, a series of panels show chronological evolution of the formation and growth of charge-clusters for time intervals of 25, 55, 100, and 150 unit time as described in the figure caption. The formation of a charge-cluster is defined by the emergence of a region with $c_{(\bf r,t)} >$ 0.7. These results highlight the dependence of the population and size distribution of charge-clusters overtime on the nucleation rate $R$ through $P_{n}$, a larger $R$ results in higher number density and smaller average size. 

In the next section, we describe how the formation and subsequent growth of charge-clusters lead to the retention loss for both OFF-state and ON-state.

\section{\label{sec:level3}THE EMERGENCE OF CHARGE-CLUSTERS AND THE RETENTION LOSS }

In our previous study, cyclic switching through the formation and annihilation of ECFs in a dielectric film – switching layer – is depicted by coupling electrical and thermal transport using the Cahn-Hilliard phase-field model\cite{t19}. The study highlighted how the formation and evolution of charge-clusters within a switching layer are driven under the influence of electric potential applied across the layer. The study also elucidated that experimentally obtained $R_{OFF}$ is always lower than that of the pristine state (i.e., the state established in as-fabricated memristors before a necessary conditioning often referred to as electroforming is performed). $R_{OFF}$ established during a RESET operation is dominated by the gap between one of the two electrodes and the tip of an ECF located closest to the electrode; thus, the highest electric-field $E_{gap}$ is expected to appear over the gap during the rest of the RESET operation and the subsequent SET operation. It is this $E_{gap}$ that would initiate the formation of charge-clusters as discussed in Section I. The formation of charge-clusters is illustrated in the view of the classical nucleation theory that encompasses an initial slow nucleation stage and a subsequent fast nucleation stage before the coalescence of nuclei (i.e., the formation of charge-clusters). 

\begin{figure}[ht]
	\centering
	\scalebox{.42}{\includegraphics{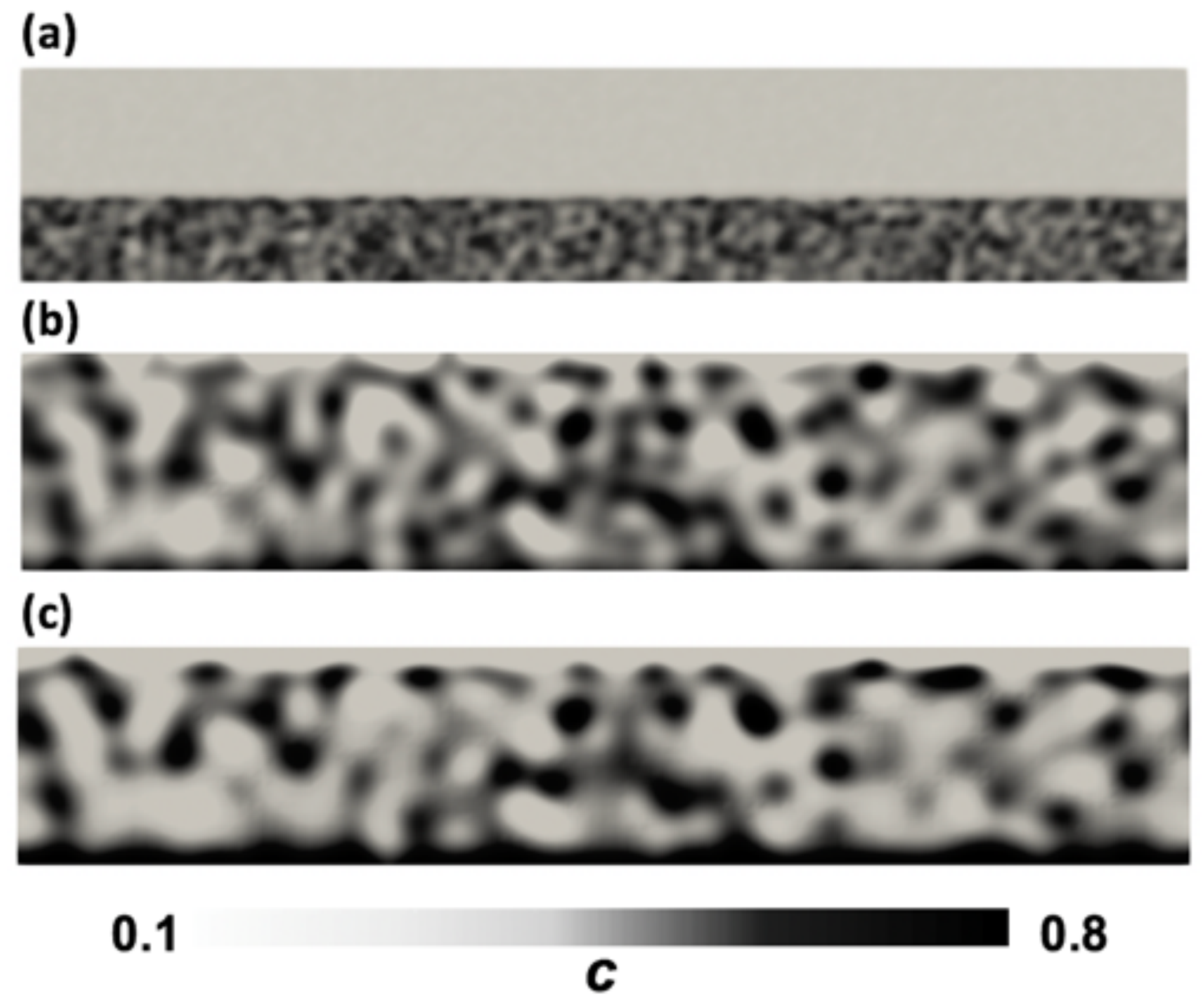}}
	\caption{A switching layer with dimension of 50$nm$ x 10$nm$ set at temperature of 400 $K$. (a) pristine-state, (b) ON-state established by applying electrical potential of 1$V$ to the upper bound of the layer and (c) OFF-state obtained subsequently by reversing the polarity of the electrical potential}
	\label{fig:nuc_two}
\end{figure}

Fig.~\ref{fig:nuc_two}(a) shows a system made of a 50 $nm$ x 10 $nm$ switching layer in its pristine state. The system consists of two distinctive regions: the lower region being electrically conducting and the upper region being electrically insulating. The conducting region is represented by varying contrast that signifies local variations in $c_{(\bf r,t)}$. The insulating region shows uniform $c_{(\bf r,t)}$ set to zero. The conducting region is distinctly separated from the insulating region by an interface along which $c_{(\bf r,t)}$ varies. At $t$ = 0 , the variations in $c_{(\bf r,t)}$ in the conducting region was produced by choosing a random number in the range of 0.7 ${\bf <}c_{(\bf r,0)}{\bf <}$ 0.9 while, in the insulating region, choices for $c_{(\bf r,0)}$ were randomly made in the range of 0.1 ${\bf <}c_{(\bf r,0)}{\bf <}$ 0.3. Fig.~\ref{fig:nuc_two}(b) represents ON-state established by applying electrical potential of 1 $V$ to the top electrode (i.e., the upper bound of the insulating region in Fig.~\ref{fig:nuc_two}(a)) at room temperature. The presence of multiple ECFs connecting the top electrode and the bottom electrode is readily identified. Subsequently, the polarity of the electrical potential was reversed to obtain OFF-state as show in Fig.~\ref{fig:nuc_two}(c), which shows that the ECFs that existed in ON-state were ruptured, leaving an insulating gap clearly visible below the top boundary. 

As described in the previous section, locations, within a switching layer, that experience electric-field locally much higher than the nominal electric filed (i.e., the applied electric potential divided by the physical thickness of the switching layer) – hot spots – during cyclic switching operations are most likely to be preferential nucleation cites for charge-clusters. In Fig.~\ref{fig:nuc_two}(b) and (c), such hot spots are identified as gaps in lighter contrast between neighboring conductive regions. With the system being in OFF-state, a SET operation causes hot spots to appear where nuclei – the precursors of charge-clusters – form and remain, eventually leading to quasi-ON-state as a result of the unintended formation of ECFs. On the other hand, with the system being in ON-state, a RESET operation causes the annihilation of ECFs, which subsequently generates gaps between remnants of ECFs; these gaps act as hot spots during a subsequent SET operation. In a microscopic view, the spatial distribution of gaps/hot spots is expected to vary from time to time because of the random nature of the formation and growth of charge-clusters. 
\\
In our assertion, the degree at which ON-state and OFF-state remain stable depends on how charge-clusters form, evolve, and rearrange themselves, as the number of cyclic switching operations (i.e., accumulative SET and RESET operations) grows. Self-diffusion of charge-clusters and/or fractions of charge-clusters is expected to govern the rearrangement of charge-clusters when no electric potential operates (i.e., a period of time between a SET and a RESET operations), and thus, locations and the number of nuclei that form during a SET and a RESET operations are expected to influence the data retention. In our simulation, nucleation was induced by introducing a free energy penalty to avoid the need for a direct modification of local $c_{(\bf r,t)}$, and instead, by modifying local energy density in such a way that a local minima of free energy density is allocated to an intended nucleation site, forcing the neighboring charge-clusters to diffuse toward the nucleon site. This approach allows us to write the total homogeneous free energy density of the system, $f_{T}$, to be comprised of bulk free energy density of as two-phase system – a system consists of electrically conducting and non-conducting regions in our case – \cite{t19} and nucleation free energy density $f_{n}$:
\begin{eqnarray}
f_{T}(c,T)=f_{{bulk}}(c,T) + f_{n}(c,T)
\label{eq:nuc_five}
\end{eqnarray}

where $f_{{bulk}}(c,T)$ is the temperature dependent double-well free-energy density function defined in the diffuse interface approximation to suitably describe dynamical structural evolution of the system, which was introduced in our previous work\cite{t19} and expressed as:
\begin{eqnarray}
f_{{bulk}}(c_{(\bf r,t)},T)= A\left[c_{(\bf r,t)}-c_{1}\right]^2 \left[c_{(\bf r,t)}-c_{2}\right]^2 \left(1- \frac{T}{T_{c}}\right)^n
\label{eq:nuc_six}
\end{eqnarray}
where $A$ is the magnitude of the double-well potential, $c_{1}$ and $c_{2}$ represent the normalized charge concentration of conducting and non-conducting states, respectively, $T_{c}$ is the critical temperature of the system, and $n$ is an empirical factor and it was assumed to be 2 in our simulations \cite{t19}. $c_{(\bf r,t)}$ varies within the range of 0 ${\bf <}c_{(\bf r,0)}{\bf <}$1.

$f_{T}$ defined in Eq.~(\ref{eq:nuc_five})  is used in the phase-field method to calculate the total free energy of the system $f(c,T)$ that needs to be minimized:
\begin{eqnarray}
F_{(c,T)}=\int_{S}[f_{T}(c_{({\bf r},t)},T)+\frac{1}{2}\kappa |  \nabla c_{({\bf r},t)}|^2]ds
\label{eq:nuc_seven}
\end{eqnarray}
where the integration is done for the entire system represented by $S$. The dynamical evolution of ECFs originated at the interface formed between a conducting and nonconducting regions is described by modified Cahn-Hilliard equation:
\begin{eqnarray}
\frac{\partial c_{({\bf r},t)}}{\partial t}= \nabla . (M  \nabla \frac{\partial {F_{(c,T)}}}{\partial c_{({\bf r},t)}})
\label{eq:nuc_eight}
\end{eqnarray}
\begin{eqnarray}
\frac{\partial c_{({\bf r},t)}}{\partial t}= \nabla . M  \nabla \left[ \frac {\partial {f_{T}(c,T)}}{\partial c_{({\bf r},t)}} - \nabla. \kappa  \nabla c_{({\bf r},t)} \right]
\label{eq:nuc_nine}
\end{eqnarray}
where $M$ is the mobility of the conserved variable $c_{({\bf r},t)}$, which is assumed to be constant in our simulations. In order to introduce the nucleation of charge-clusters in the systems prepared in Fig.~\ref{fig:nuc_two}(b) and (c) representing ON-state and OFF-state, respectively, the following method was used to find specific locations at which the nucleation was most likely to take place. 

\begin{figure}[ht]
	\centering
	\scalebox{0.8}{\includegraphics{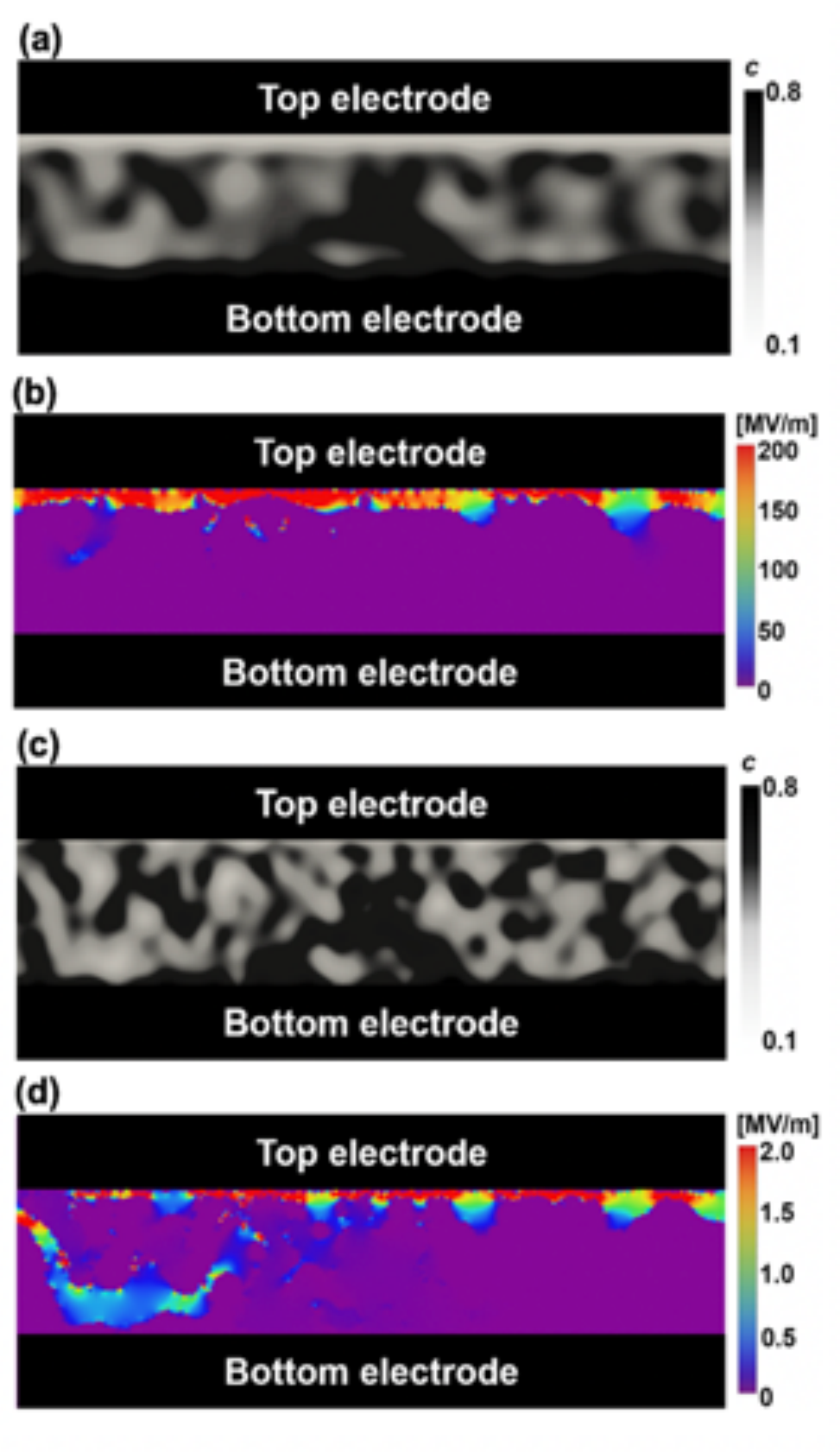}}
	\caption{(a) a charge concentration map of the system in OFF-state, highlighting the presence of a nearly continuous non-conducting region at the top interface, (i.e. light gray region). (b) an electric-field map associated with the charge concentration map in (a). Regions experiencing high electric-field (i.e., regions in red) are located underneath the top electrode. (c) a charge concentration map of system in ON-state, after formation of a few ECFs (i.e., continuous dark regions) connecting the top electrode to the bottom electrode (d) an electric-field map of the system in panel (c). Regions experiencing high electric-field appear beneath the top electrode; however, its magnitude is much lower than that in (b).}
	\label{fig:nuc_three}
\end{figure}

As explained in the previous section, since the probability of nucleation is expected to be high in regions locally experiencing high electric-field, electric-field maps were generated for systems in different states (i.e., either OFF-state or ON-state) to identify a potential nucleation sites in these two states. In order to generate electric-field maps, a pair of 5 $nm$ thick layers with c uniformly set to 1 was added to the concentration maps that represent OFF-state and ON-state to provide a top and a bottom electrode, as shown in Fig.~\ref{fig:nuc_three}(a) and (c), respectively. Then, an electric potential of 1 $V$ was applied to the top electrode while the bottom electrode was grounded. Resulting maps of electric-field for OFF-state and ON-state are shown in Fig.~\ref{fig:nuc_three}(b) and (d), respectively. 

In Fig.~\ref{fig:nuc_three}(b), regions locally experiencing electric-field higher (i.e., regions in red) than that of the majority region in purple (i.e., electric-field is near zero) are, as expected, mainly located within the narrow gap separating the top electrode from the uppermost conducting region. This is plausible even for ON-state because, in a filamentary memristor, an applied electric potential in a system in ON-state (i.e., a system that contains ECFs) drops across the ECFs during a SET operation, and thus, non-conducting regions still experience relatively high electric-field potentially similar to that evolves during a RESET operation. Our assessment suggests that, during a RESET operation, local electric-field is higher than that arises during a SET operation (i.e., an operation that establishes ECFs) by a few orders of magnitude inarguably because ECFs fractur during a RESET operation, which highlights the fact that the probability of nucleation during a RESET operation is higher than that during a SET operation.
\\
Once specific regions in which higher local electric-field is most likely to develop for the systems in OFF-state (Fig.~\ref{fig:nuc_three}(b)) and in ON-state (Fig.~\ref{fig:nuc_three}(d)) were identified, then the introduction of nuclei that grow into charge-clusters was carried out to study impacts of the nucleation.

\begin{figure}[ht]
	\centering
	\scalebox{0.42}{\includegraphics{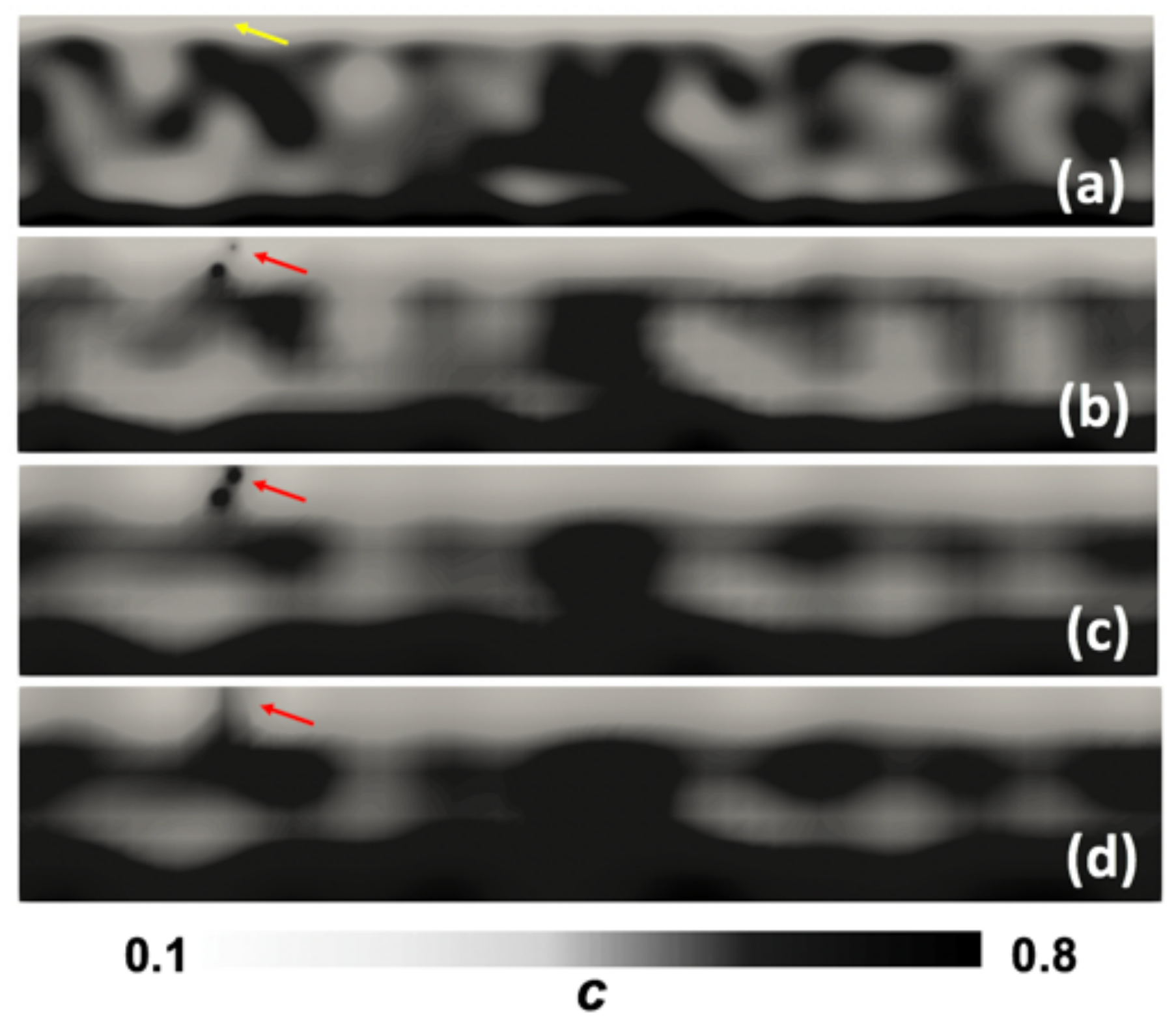}}
	\caption{Formation and growth of charge-clusters in the system in OFF-state at $T$ = 400 $K$. (a) the system was prepared to be in OFF-state, after completing five SET/RESET cyclic switching operations, showing the presence of a continuous non-conducting region beneath the top boundary. The yellow arrow points the nucleation cite at which two nuclei would be introduced in (b). (b ) the system left in OFF-state with two nuclei introduced at the selected nucleation cite, representing the system being at $t$ = 0 unit time after the removal of the last RESET bias, (c) the nuclei grew into a charge-cluster and eventually merging into a fractured ECF at $t$ = 40 unit time, and then, (d) a complete ECF was formed, which results in the retention loss of the system in OFF-state, at $t$ = 100 unit time. }
	\label{fig:nuc_four}
\end{figure}

\begin{figure}[ht]
	\centering
	\scalebox{0.6}{\includegraphics{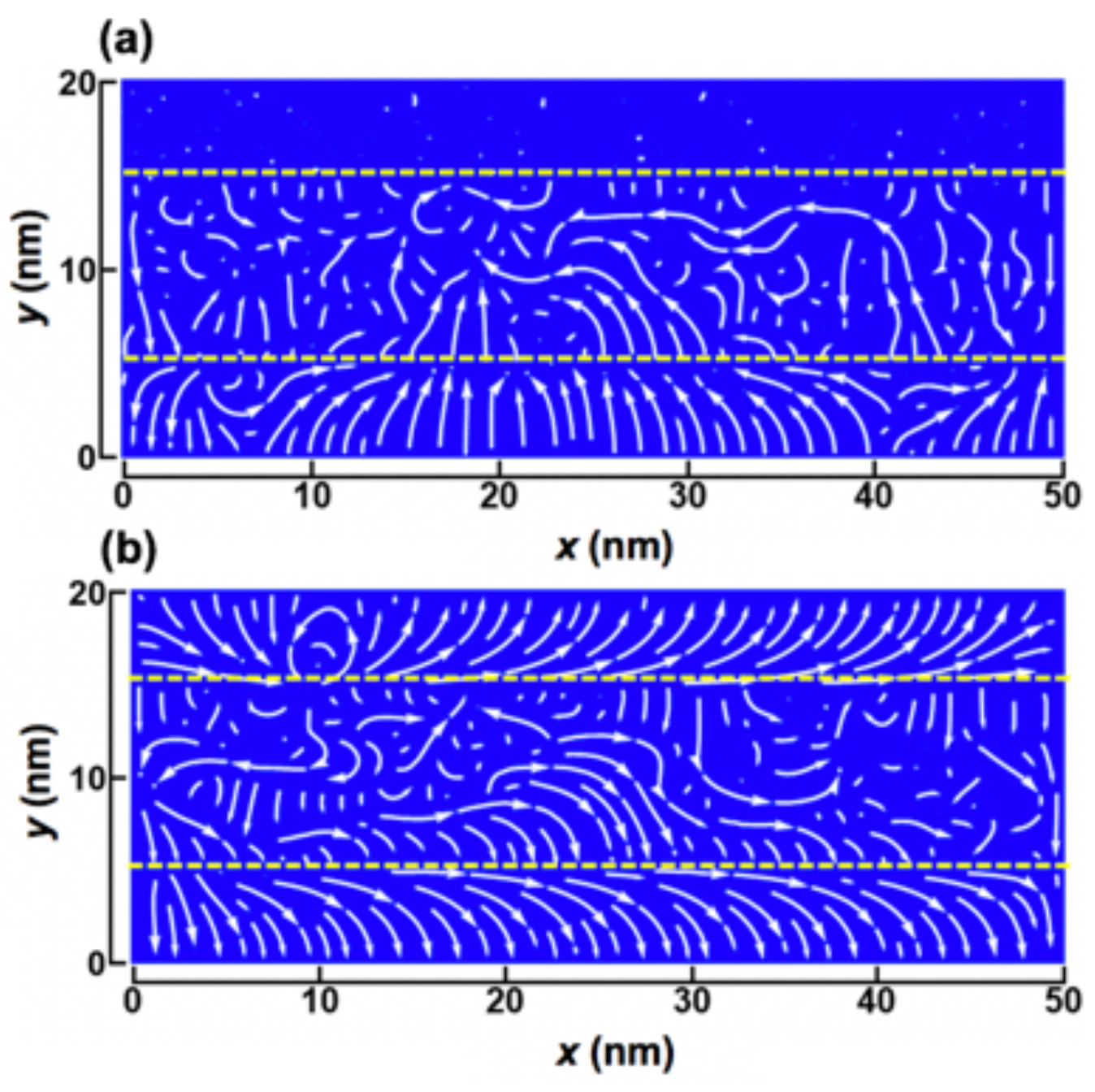}}
	\caption{A current density map of the system in OFF-state (a) before and (b)after the introduction of the nuclei. (a) before the introduction of nuclei, $j_{rel}(r)$ vanishes in the top electrode because the system is set to OFF-state, (b) the introduction and growth of  the nuclei leads to the formation of an infelicitous ECF, which results in $j_{rel}(r)$ being continuous from the top to the bottom electrode, indicating the presence of a ECF as seen in Fig.~\ref{fig:nuc_four}(d); thus, the OFF-state established in panel (a) is destroyed and the system is in an erroneous OFF-state.}
	\label{fig:nuc_five}
\end{figure}

To prepare a system in OFF-state for the introduction of nuclei, the charge-cluster concentration map shown in Fig.~\ref{fig:nuc_four}(a) was established by completing four additional SET/RESET cycles after the first SET/RESET cyclic switching operation was completed in Fig.~\ref{fig:nuc_two}(c). The impact of multiple SET/RESET cyclic switching operations is pronounced in the overall blurriness of charge-clusters seen in Fig.~\ref{fig:nuc_four}(a) in comparison to Fig.~\ref{fig:nuc_two}(c). The yellow arrow in Fig.~\ref{fig:nuc_four}(a) points to a specific location within the non-conducing gap separating the top electrode from the network of conductive regions connected to the bottom electrode. This specific location was selected as a representative nucleation site experiencing higher electric-field during SER/RESET cyclic switching operations. In Fig.~\ref{fig:nuc_four}(b), two nucleus were introduced, as indicated by the red arrow, at the selected nucleation site. The number of nuclei added to the system was assumed to be related to the cumulative electrical stress (i.e., the total number of cyclic switching operations) the system has experienced. For this study, two nuclei were introduced, as an example, for a system that has experienced multiple cyclic switching operations. Introducing larger or fewer number of nuclei should not change general conclusions derived from our study. Fig.~\ref{fig:nuc_four}(b) essentially represents a system in OFF-state at the end of the last RESET operation with the addition of two nuclei introduced as a result of high electric-field, thus, Fig.~\ref{fig:nuc_four}(b) defines $t$ = 0 at which the last RESET bias was removed. Fig.~\ref{fig:nuc_four}(c) and (d) illustrate the growth of these nuclei over time, at $t$ = 40 and $t$= 100 unit time, respectively; the nuclei eventually merged and grew, connecting the top electrode to the upper portion of the network of conductive regions as the free energy of the system was reduced. Using the $c(\bf r)$ maps in Fig.~\ref{fig:nuc_four}(b) and (d), maps of relative electrical current density $j_rel(r)$ were obtained after addition of a top and a bottom electrode across the system, as explained before, and are illustrated in Fig.~\ref{fig:nuc_five}(a) and (b) where the black dashed lines represent the interface between the electrodes and the switching layer. The $j_{rel}(r)$ maps were produced during the READ operation, by applying an electric potential of $V_{top}$ = 100 $mV$ to the top electrode while the bottom electrode was grounded. The magnitude and direction of current density at a specific location is expressed by the length and direction of a white arrow. Lengths of white arrows are relative only within a map. In Fig.~\ref{fig:nuc_five}(a) and (b), current density represented by white arrows is overlaid on a map of associated electric-field. Relative electrical current Irel calculated by integrating current density along the width (i.e., 0-50 $nm$) of the switching layer is $I_{rel}(r)$ = 2x10-6 and $I_{rel}(r)$ = 7 for Fig.~\ref{fig:nuc_four}(a) and (b), respectively. The $j_{rel}(r)$ maps clearly highlight that, in a system left in OFF-state after the removal of the last RESET bias, ECFs could form as a result of the emergence of charge-clusters originated to nuclei centers produced during the previous SET/RESET cyclic switching operations as the free energy of the system reduces, which results in the retention loss of the system in OFF-state.

\begin{figure}[ht]
	\centering
	\scalebox{0.42}{\includegraphics{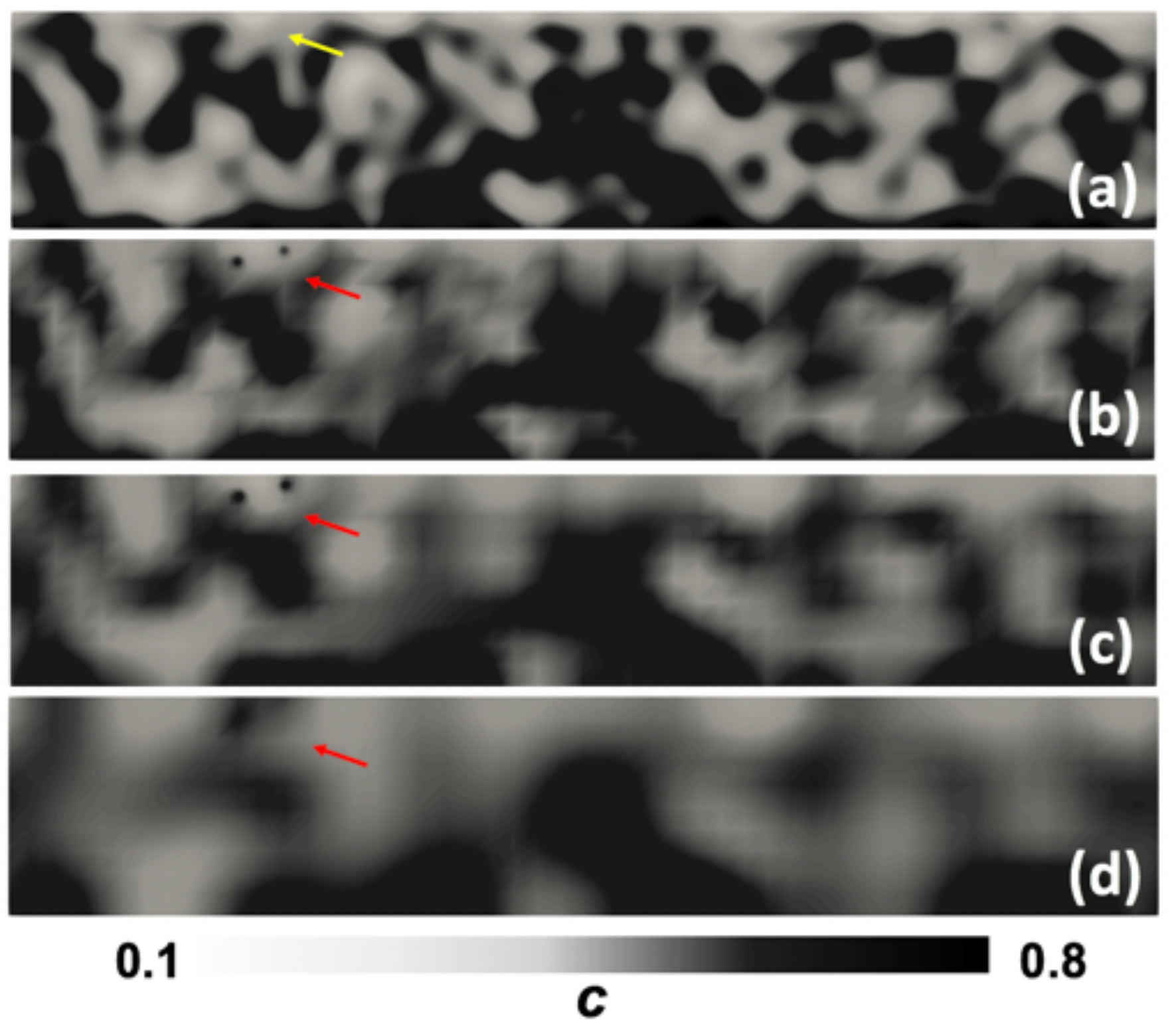}}
	\caption{Generation and growth of charge-clusters in the system in ON-state at $T$ = 400 $K$. (a) the system was prepared to be in ON-state after completing five SET/RESET cyclic switching operations and an additional SET operation, resulting in the formation of multiple ECFs connecting the top and the bottom electrodes. The dominant ECF is located at the center. The yellow arrow pointing to a representative region that experienced high electric-field during the SET/RESET cycles was selected for the introduction of nuclei, (b ) generation of two nuclei within the gap in the top-left side of the system indicated by the red arrow at $t$ = 0 unit time, (c) at $t$ = 40 unit time and (d) at $t$ =100 unit time show continuous growth of the nuclei into aggregated charge-clusters and into a ECF, eventually forming a new dominant ECF.}
	\label{fig:nuc_six}
\end{figure}

\begin{figure}[ht]
	\centering
	\scalebox{0.6}{\includegraphics{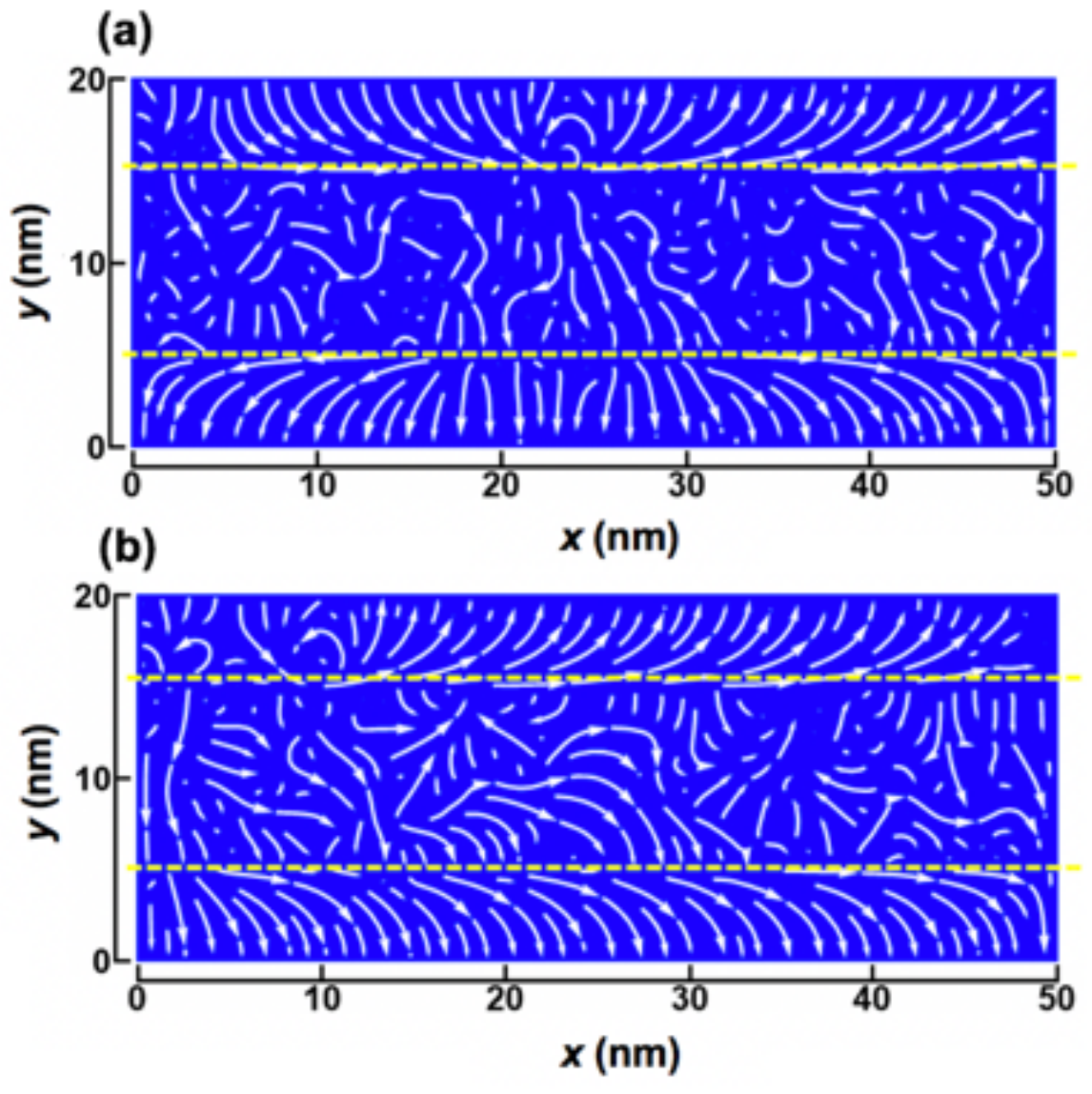}}
	\caption{A current density map of the system in ON-state (a) before and (b)after the introduction of the nuclei. (a) a $j_{rel}(r)$ map shows the presence of the dominant ECF at the center connecting the top electrode to the bottom electrode as the system is in ON-state as established in Fig.~\ref{fig:nuc_six}(a), (b) a $j_{rel}(r)$ map for Fig. 6(d) in which characteristics (i.e., spatial distribution and density)of white arrows appear to be completely different from what is seen in (a), illustrating distinctive changes that occurred in the system as a result of the introduction and growth of the nuclei even without the influence of an external electrical potential.}
	\label{fig:nuc_seven}
\end{figure}

Impacts of the nucleation in a system left in ON-state were also investigated. Fig.~\ref{fig:nuc_six}(a) shows a exemplifying system in ON-state prepared after completing five SET/RESET cyclic switching operations. Adopting the same approach as explained for the OFF-state examined above, a representative of regions experiencing high electric-field during the last SET/RESET operation was identified as a nucleation site and indicated by the yellow arrow in Fig.~\ref{fig:nuc_six}(a). Considering Fig.~\ref{fig:nuc_six}(b) being at $t$ = 0 unit time, two nuclei were introduced after the last SET bias was removed and indicated by the red arrow in Fig.~\ref{fig:nuc_six}(b). Fig.~\ref{fig:nuc_six}(c) and (d) illustrate the growth of these nuclei over time, at $t$ = 40  and $t$ = 100 unit time, respectively. The results clearly show that the nuclei eventually merged and grew into a fractured ECF as the total free energy of the system was reduced, and the dominant ECF located approximately at the center (i.e., x ~25 $nm$) of the figure appeared to be less pronounced or even destroyed. For a better illustration, current density maps associated with Fig.~\ref{fig:nuc_six}(b) and (d) are produced in the same way as that described for Fig.~\ref{fig:nuc_five}(a) and (b), and are shown in Fig.~\ref{fig:nuc_seven}(a) and (b), respectively. A comparison between Fig.~\ref{fig:nuc_seven}(a) and (b) clearly shows that the dominant ECF originally present in the center of Fig.~\ref{fig:nuc_seven}(a) disappears in Fig.~\ref{fig:nuc_seven}(b), instead, two new ECFs are formed; one roughly at the location of nuclei centers  (i.e., x ~10 $nm$) and the other at a location (i.e., x ~3 $nm$) that does not appear to be directly related to the nucleation but rather the growth of a fractured ECF remnant from the previous RESET operation. Calculated $I_{rel}(r)$ is 12 for Fig.~\ref{fig:nuc_seven}(a) and 24 for Fig.~\ref{fig:nuc_seven}(b), that is, $R_{ON}$ is smaller for Fig.~\ref{fig:nuc_seven}(b) compared to Fig.~\ref{fig:nuc_seven}(a) – the establishment of deeper ON-state. Even though the retention was not lost for this particular system originally set to On-state, an increase in the conductance would potentially lead to an irreversible failure during the subsequent RESET operation as more electric power has to be delivered to the system in order to disconnect the ECFs, which consequently could also increase cycle to cycle variabilities if the RESET operation fails to complete fully. All these outcomes would eventually contribute to the overall reliability. 

\begin{figure}[ht]
	\scalebox{0.48}{\includegraphics{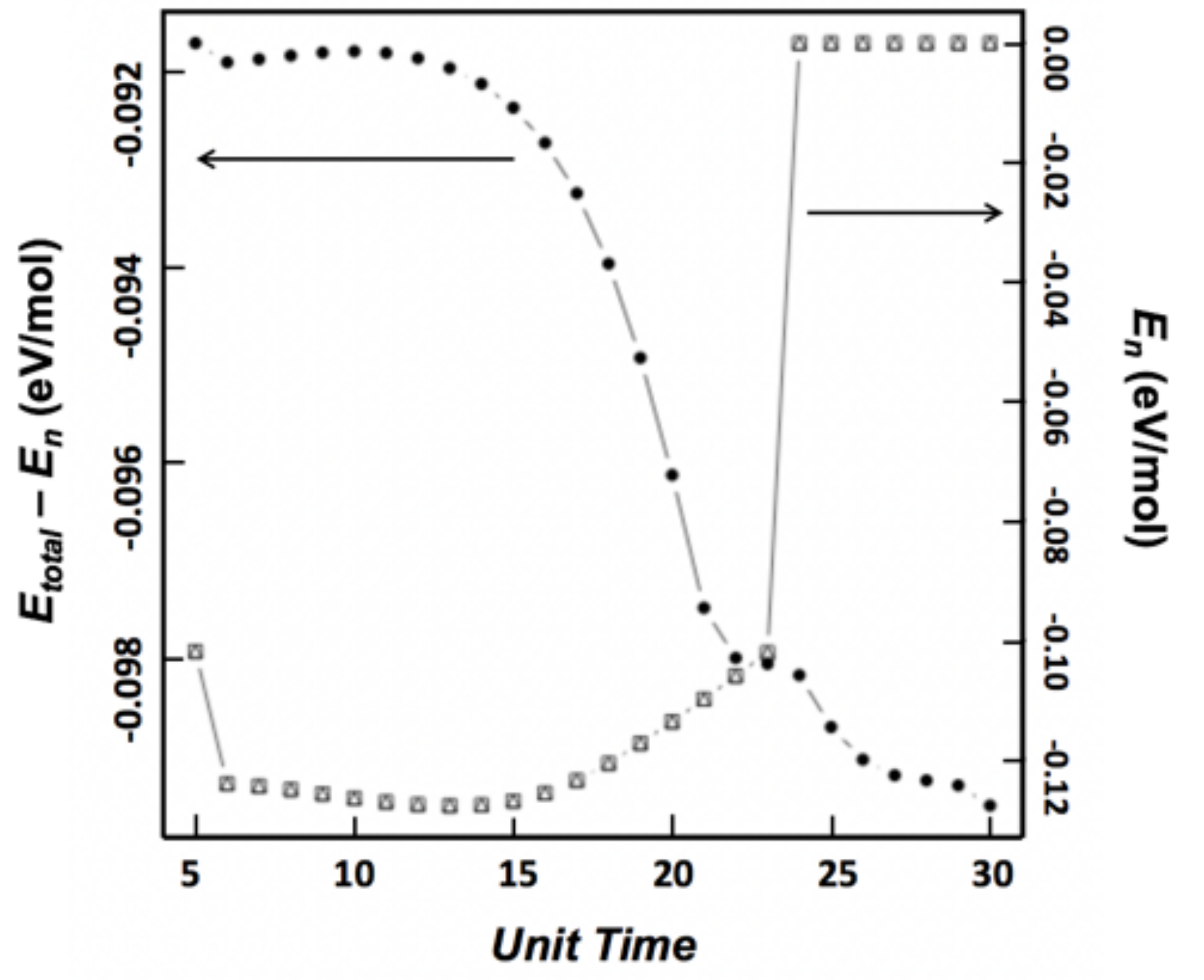}}
	\caption{Evolution of $E_{total} - E_{n}$ and $E_{n}$ during the first 30 unit time of the simulation. A sharp transition occurs in $E_{total} - E_{n}$ as $E_{n}$ approaches to zero at around 23 unit time.}
	\label{fig:nuc_eight}
\end{figure}

Fig.~\ref{fig:nuc_eight} illustrates the evolution of the nucleation energy density $E_{n}$ and the difference between the total free energy density $E_{total}$ and $E_{n}$ (i.e., $E_{total} - E_{n}$); both $E_{n}$ and $E_{total} - E_{n}$ were obtained by averaging their respective local values within a portion – 2 $nm$ thick – of the switching layer underneath the top electrode where nucleation centers were introduced (i.e., the region that experienced local electric-field higher than the nominal electric filed during as seen in Fig.~\ref{fig:nuc_three}). Fig.~\ref{fig:nuc_eight} indicates that the rate at which $E_{total} - E_{n}$ is lowered is highly impacted by changes in $E_{n}$. The sharp transition in $E_{total} - E_{n}$ begins at unit time 17 as $E_{n}$ starts approaching to zero suggesting that the nucleation would directly impact local rearrangements of charged clusters and influence the short-term reliability. 

The nucleation not only impacts the eventual evolution of charged clusters within a system but also it plays an important role in processes by which a switching layer lowers its total free energy over time; in another way of saying, the nucleation is expected to influence both short-term as well as long-term reliability.

\section{\label{sec:level5}Summary }
In this study, we introduced the nucleation of charge-clusters as a potential source to address a critical failure mode – the retention loss – often experimentally observed in operating memristors. We analyzed impacts of the nucleation of charges and the growth of nuclei into charge-clusters in systems made of a dielectric layer that represents a switching layer in memristors. We separately analyzed systems set to either OFF-state or ON-state. We employed the well-known phenomena – high electric-field promotes nucleation – and to determine possible sites for the introduction of nuclei in switching layers made of a dielectric thin-film that underwent a high electric stress during multiple SET/RESET cyclic switching operations. Our study demonstrated that nuclei introduced in a switching layer grow dynamically, potentially change the state of the device when they grow into charge-clusters, and further develop structural variations that continuously evolve over time even without an external electric potential. More specifically, in a system set to OFF-state, the growth of nuclei can eventually create ECFs, resulting in the retention loss. In contrast, in a system set to ON-state, the growth of nuclei can potentially complete random fractured ECFs and substitute the dominant ECF, eventually contributing to device-to-device and cycle-to-cycle variabilities as electrical power required for the completion of the subsequent RESET operation increases, resulting in long-term reliability degradation of the system. Furthermore, our study showed that the nucleation can influence the rate at which local total free energy density reduces, highlighting its impact on both short-term and long-term reliability.
The data that support the findings of this study are available from the corresponding author, upon reasonable request.

\section*{References}
\bibliography{nuc_paper}
\end{document}